# A Physics-Constrained Learning Framework for Wave Propagation in Complex Poroelastic Multilayered Media

Ya Gao[1, 2, 3], Yifan Wang[1], Yiming Chen[1], Haohan Sun[1], Shoukun Lyu[1], Junmei Cao[1], Weijiang Xu[4, 5*], Qian Cheng[1, 2, 3*]

[1]Institute of Acoustics, School of Physics Science and Engineering, Tongji University, Shanghai 200092, China.

[2]Department of Ultrasonography, Shanghai Disabled Persons' Federation Key Laboratory of Intelligent Rehabilitation Assistive Devices and Technologies, Yangzhi Rehabilitation Hospital (Shanghai Sunshine Rehabilitation Center), Tongji University School of Medicine, Shanghai 201619, China.

[3]National Key Laboratory of Autonomous Intelligent Unmanned Systems, Shanghai Research Institute for Intelligent Autonomous Systems, Tongji University, Shanghai 201210, China.

[4]CNRS UMR 8520 IEMN, Université Polytechnique Hauts-de-France, Site de Valenciennes, F-59313 Valenciennes, France.

[5]INSA Hauts-de-France, F-59313 Valenciennes, France.

[*]Corresponding author E-mail: q.cheng@tongji.edu.cn, wei-jiang.xu@uphf.fr

## Abstract

Wave propagation through complex poroelastic multilayered media is difficult to model and invert because pronounced heterogeneity, scattering, mode conversion and fluid–solid coupling jointly distort acoustic signals during propagation. Here we present Physics-Constrained Learning for Complex Multilayered Media (PCL-CMM), a general framework that integrates Biot's poroelastic theory with the elastic wave equation to bridge the gap between physically rigorous wave modelling and data-driven learning. PCL-CMM constructs a high-fidelity digital twin that dynamically computes an effective acoustic stiffness tensor for forward wave modelling and incorporates the resulting physical constraint as a loss term to regularize the training of deep neural networks. We demonstrate PCL-CMM on transcranial photoacoustic imaging, where skull-induced acoustic distortions severely degrade image formation. Across simulations and ex vivo experiments, PCL-CMM effectively compensates for these distortions and improves SSIM by more than 0.06 compared with purely data-driven neural networks. This work establishes a physics-constrained learning framework for acoustic wave modelling in complex poroelastic multilayered media.

## Main

Accurately modelling and interpreting wave propagation in complex media remains a fundamental and long-standing challenge across various fields of science and engineering that rely on wave phenomena. systems such as biological tissues[1–3], geological formations[4–6], and composite materials[7,8] often exhibit strong heterogeneity, anisotropy, and significant acoustic impedance mismatches—especially in the presence of porous media layers—leading to complex scattering, mode coupling, and energy dissipation[9–11]. Such complexity renders analytical solution of the wave propagation governing equation intractable, while simultaneously amplifying both parametric uncertainty and computational burden in numerical forward modeling and inverse reconstructions.

To describe these complex processes, researchers have developed a wide range of approaches, including time-reversal[12–14], wavefront shaping[15–17], transmission matrix[18–20], full-wavefield modeling[21–23], and machine-learning-based modelling strategies[24,25]. These methods have achieved notable progress under specific conditions and have successfully captured the spatial variability and energy distribution of wave propagation. However, in multilayered complex media containing porous materials, several key challenges remain: (i) the sharp interlayer discontinuities between layers and the coupling of acoustic parameters strongly influence the phase continuity and energy redistribution of the wavefield; (ii) the microscopic porosity and viscoelasticity determine the frequency-dependent macroscopic response; and (iii) achieving an effective balance between computational feasibility and high physical fidelity in multiscale coupled systems remains an open problem.

To address these challenges, we propose a unified framework that integrates physical modelling with data-driven learning, termed Physics-Constrained Learning for Complex Multilayered Media (PCL-CMM). This framework incorporates physical laws as soft-coded physical constraints, tightly coupling the Biot's poroelastic theory with the elastic wave equation, thereby establishing a high-fidelity digital twin that synergizes the actual acoustic fields with the underlying physical fields, enabling the precise capture of multiphysics interactions within complex layered structures. A dynamically generated acoustic stiffness tensor serves as a physics-informed prior which effectively guides the training of deep neural networks toward physically consistent inverse solutions. To validate its efficacy, we apply the proposed framework to one of the most challenging inverse problems in transcranial photoacoustic imaging (tPAI) area. In this scenario, the skull's inherent heterogeneity and multilayered anatomical architecture induce substantial acoustic aberrations and waveform distortions, posing a fundamental limitation to high-resolution deep-brain tPAI. By synergistically integrating digital-twin modelling with physics-informed learning, PCL-CMM enables accurate characterization and compensation of skull-induced wavefield distortions, facilitating high-fidelity reconstruction of brain structural signals. This work establishes a generalizable and robust paradigm for physics-consistent wavefield reconstruction and inverse modeling in complex multilayered environments, with broad implications for advancing non-invasive neuromodulation and related fields.

## Results

### PCL-CMM: a physics-constrained learning framework for complex multilayered media

A unified framework is developed to characterize wave propagation in complex multilayered media across different disciplines, serving as the foundation of the PCL-CMM methodology (Fig. 1). In biomedical imaging, geophysical exploration and composite materials design, structurally diverse media can be abstracted at the macroscopic level as stacks of fluid, dense elastic solid and porous solid layers (Fig. 1a). Within this unified modeling framework, wave physics is rigorously represented: longitudinal waves in fluids, longitudinal and transverse waves in elastic solids, and fast/slow longitudinal and transverse waves in porous layers based on Biot's theory (Fig. 1b). This physically grounded, cross-domain abstraction facilitates a consistent, mechanism-aware description of multimodal wave propagation across these ostensibly distinct application domains.

For each of these three representative layer types, layer-specific characteristic matrices and stiffness matrices are systematically derived and recursively assembled into a global stiffness matrix. This assembly strictly enforces displacement and stress continuity across all internal interfaces (Fig. 1c), yielding a frequency- and angle-resolved formulation that fully captures dispersive and anisotropic wave behavior. This resulting framework rigorously reproduces hallmark physical phenomena observed in structurally complex media, which phenomena that remain difficult to quantify with conventional approaches, including strong interfacial reflections and multi-path wave transmissions, longitudinal–transverse mode coupling, and pronounced angular and spectral sensitivity of transmission spectra.

The aggregate dynamic response of the multilayered system is subsequently encoded into a physics-constrained stiffness prior, which serves as a compact representation that encapsulates the medium's wave-modulation characteristics under arbitrary incident conditions. This stiffness prior is embedded as an explicit and soft-coded physical constraint within the architecture of a deep neural network and trained jointly with the experimental data. As a result, the framework deliver high-fidelity and physically consistent reconstructions for inverse wave propagation problems in structurally intricate multilayered media (Fig. 1c).

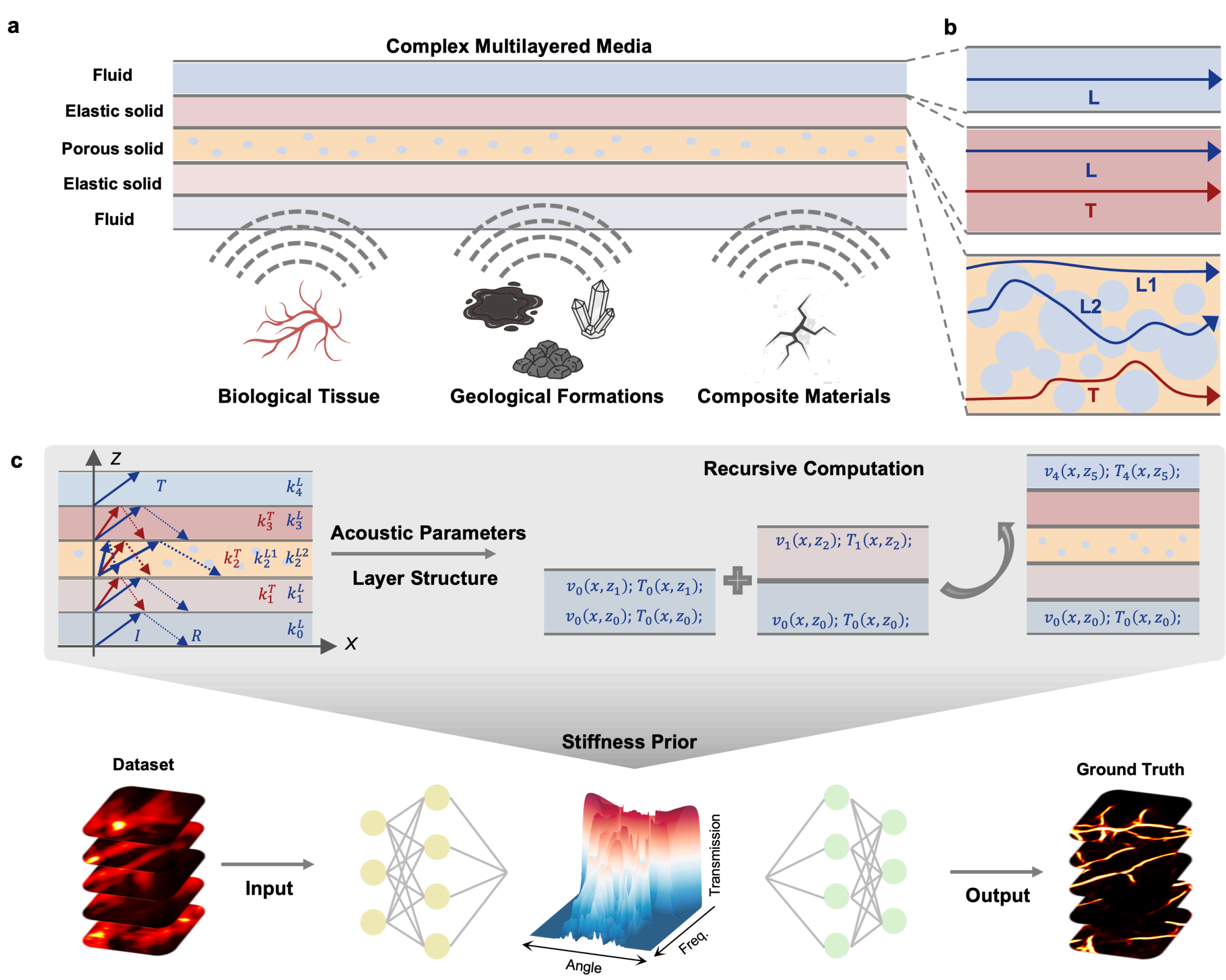

**Fig. 1. Physics-constrained learning framework for wave propagation in complex multilayered media (PCL-CMM).** a) Structurally heterogeneous multilayered media encountered in biological tissues, geological formations and composite materials can be idealized as layered stacks of fluid, elastic solid, and porous solid components. b) Within this modeling framework, fluid layers support longitudinal (L) waves, elastic solid layers support longitudinal (L) and transverse (T) waves, and porous solid layers-described by Biot's poroelastic theory-support four propagating modes: fast (L1) and slow (L2) longitudinal waves, and transverse (T) waves. c) PCL-CMM systematically formulates layer-specific characteristic and stiffness matrices for these the three representative layer types and recursively assembles them into a global stiffness matrix, which computes the transmission response as a function of incidence angle and frequency. This dynamic response is encoded into a stiffness prior that serves as a soft-coded physical constraint in a deep neural network, enabling it to learn a physically grounded mapping from experimental data to the underlying ground-truth material parameters.

**Characterization of Transcranial Wave Propagation**

We construct the global stiffness matrix $K_{G,N}$ by recursively assemble individual layer stiffness matrices using a Rokhlin-Wang-type algorithm[26], thereby rigorously defining the trancranial transmission spectrum operator $T_{skull}(\omega, k_x, D)$ with the skull thichness of $D$ (Fig. 2a). This operator is elaborated in detail within the Methods and Supplementary Notes 1 and 2. Physical priors are integrated via k-space spectral modulation: the incident acoustic pressure field $p_0(t, x, z)$ along $z$ direction is transformed into the wavenumber–frequency domain $\tilde{p}_0(\omega, k_x, z)$ via FFT; it is then modulated by $T_{skull}(\omega, k_x, D)$ to account for energy redistribution, phase delay, and mode conversion across the skull layers; finally, the modulated spectrum $\tilde{p}_{skull}(\omega, k_x, z)$ is inverse-transformed (IFFT) to reconstruct the transcranial field $p_{skull}(t, x, z)$, thereby simulating realistic structural aberrations induced by the heterogeneous skull geometry. Fig. 2b illustrates the above transformation process on the $x$-$z$ plane.

Based on the systematic characterization of the transcranial acoustic transmittance under fluid-embedded single-layer configurations (Fig. 2c), the inner and outer cortical bone layers (Layers 1 & 3) exhibit pronounced spatial and frequency selectivity -- directly attributable to their physical thickness, as unambiguously evidenced by the corresponding transmittance maps. Under near-normal incidence, transmittance is primarily governed by the frequency-thickness product; as the incidence angle approaches the first critical angle (~ 25°), a sharp transition emerges in the transmittance spectrum, signifying the onset of guided-mode excitation and interfacial wave conversion. Moreover, the substantial acoustic impedance mismatch between dense cortical bone and adjacent soft tissue results in dominant interface reflection of the incident acoustic energy, thereby severely constraining transcranial energy transmission (Fig. 2c).

In contrast, the middle diploe layer (Layer 2) displays markedly divergent wave response characteristics. Its microscopically porous structure yields reduced effective density and acoustic velocity relative to cortical bone – properties that ensure a minimal acoustic impedance mismatch between the diploe layer and soft tissue, thereby achieving a higher overall transmittance compared to the cortical bone layer. While this layer functions as an "acoustic bridge", facilitating measurable transmission even at high frequencies (3–4 MHz) and large incidence angles, its heterogeneous pore distribution induces strong acoustic scattering. This scattering degrades wavefront fidelity and constitutes the principal source of image distortion in tPAI[27].

When transitioning from isolated single layers to the full three-layer skull geometry, the integrated model accounts for the interfacial continuity between adjacent layers as well as the fluid-solid boundary conditions at the terminal incident and transmission surfaces. Crucially, constructive interference among multiple internal reflections—and coupled longitudinal-to-shear mode conversions—sustains appreciable energy transmission below ~1 MHz and within incidence angles of 30°–60°, forming a robust low-frequency transcranial acoustic window.

Quantitative characterization of one-dimensional transmittance profiles further elucidates these mechanisms. The normal-incidence (0°) curve exhibits multiple transmission peaks across the frequency band, arising from structural thickness resonances of cortical bone. At the critical angle (25°), pronounced spectral oscillations reflect complex phase accumulation and multimodal coupling. As the incidence angle further increases to 50° and 70°, transmittance decreases significantly with increasing frequency, leaving only narrowband and low-amplitude transmissions in the low-frequency band, confirming the strong angular-dependent shielding effect of the skull against oblique acoustic waves. This highly nonlinear joint dependence on frequency and incidence angle is rigorously encoded within the PCL-CMM framework and explicitly translated into structural priors for neural-network-based modeling of transcranial acoustic aberrations.

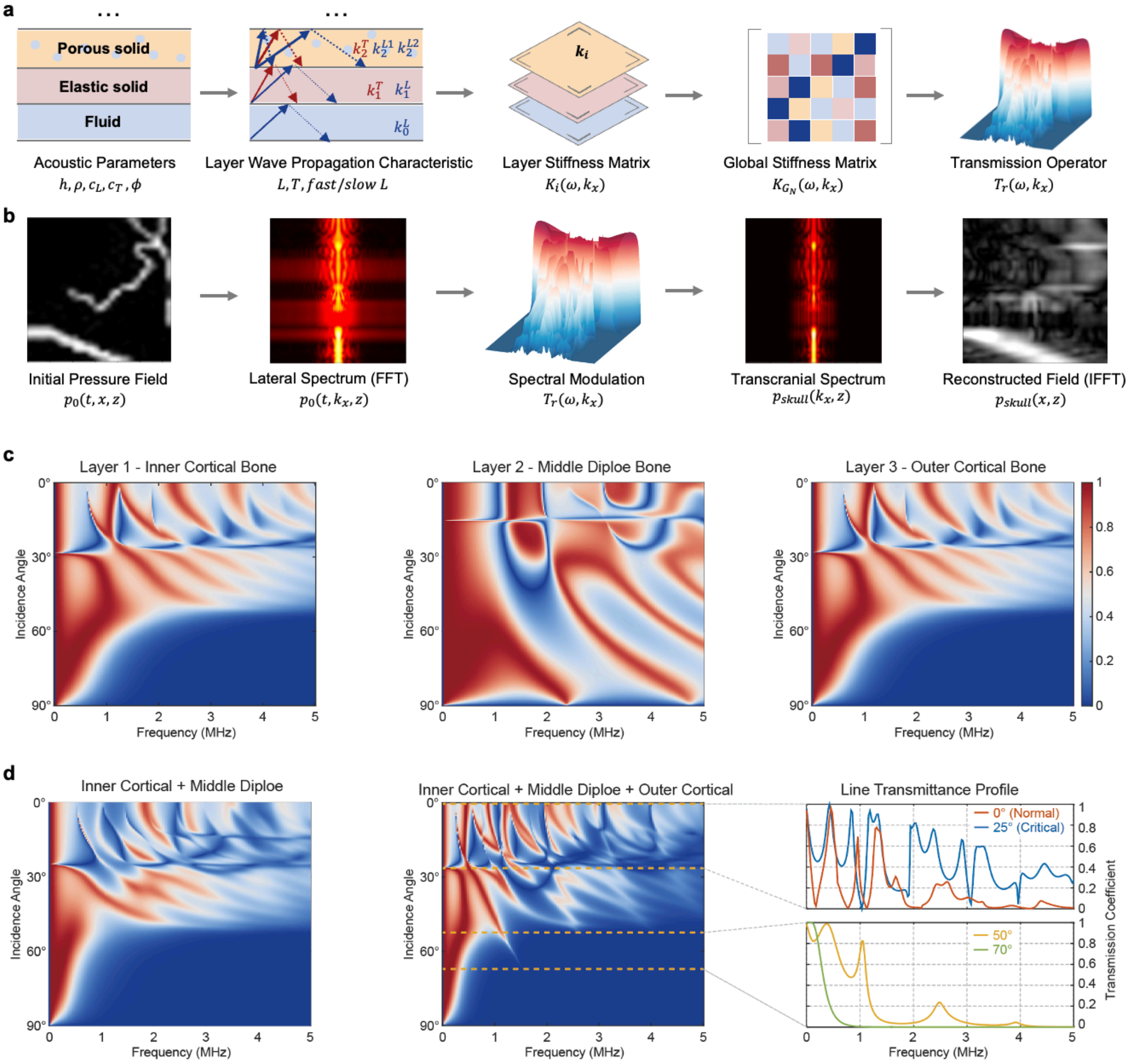

**Fig. 2. Characterization of k-space spectral modulation and transcranial transmittance.** a) Computational framework of the layered brain-skull transmission operator. b) Process of image calculation in the k-space domain. c) Transmittance spectra of isolated skull layers. Transmission coefficients for Layer 1 (inner cortical), Layer 2 (middle diploe), and Layer 3 (outer cortical) under fluid-embedded conditions. d) Transmittance of multilayered skull assemblies and angular profiles. Cumulative transmission through the two-layer (inner cortical + middle diploe) and full three-layer configurations. The included 1D line transmittance profiles specifically quantify the angular response of the full three-layer skull at representative incidence angles.

## Numerical Validation and Performance Evaluation

Numerical simulations characterize the complex propagation of photoacoustic waves as they penetrate the three-layer skull structure (Fig. 3a). Wavefield snapshots reveal strong reflection of normal stress at the skull interfaces, while shear stress exhibits significant scattering phenomena. The interaction of these multimodal waves causes severe distortion of the original wavefront before it reaches the sensor array, posing a substantial challenge for subsequent image reconstruction.

As shown in the reconstruction comparison (Fig. 3b), results using the Delay-and-Sum (DAS) algorithm relative to the vascular Ground Truth (GT) are highly blurred, with vascular structures almost entirely obscured by background clutter and artifacts. While the data-driven U-Net can recover the general outlines of the vessels, it still exhibits fractures at fine branches and retains noticeable non-physical background noise. In contrast, PCL-CMM reconstructs vascular boundaries with higher sharpness and spatial fidelity. Quantitative performance is further examined by extracting intensity profiles across various regions of the reconstructed images (Fig. 3c). The DAS profiles exhibit broad expansion and extremely low amplitudes, completely failing to localize the vascular cores. Although the U-Net curves capture the primary peaks, they show significant deviations in peak position and full-width at half-maximum. Conversely, the intensity curves of PCL-CMM nearly overlap with the GT, particularly in complex areas with multiple parallel vessels, demonstrating its ability to accurately distinguish tiny branches under the guidance of physical constraints.

The quantitative evaluation in Table 1 confirms the superior performance of PCL-CMM across all core metrics. In terms of error reduction, PCL-CMM achieves a 25.5% lower MAE compared to the data-driven U-Net. This improvement is directly reflected in enhanced image fidelity, with the framework achieving a 1.27 dB gain in PSNR and a 9.1% increase in SSIM over the purely data-driven model. These metrics indicate that by integrating the stiffness prior, the framework effectively restores complex structural details otherwise lost to skull distortions. Furthermore, a dominant 91% win rate underscores the robust consistency of the physics-constrained approach in resolving diverse anatomical aberrations.

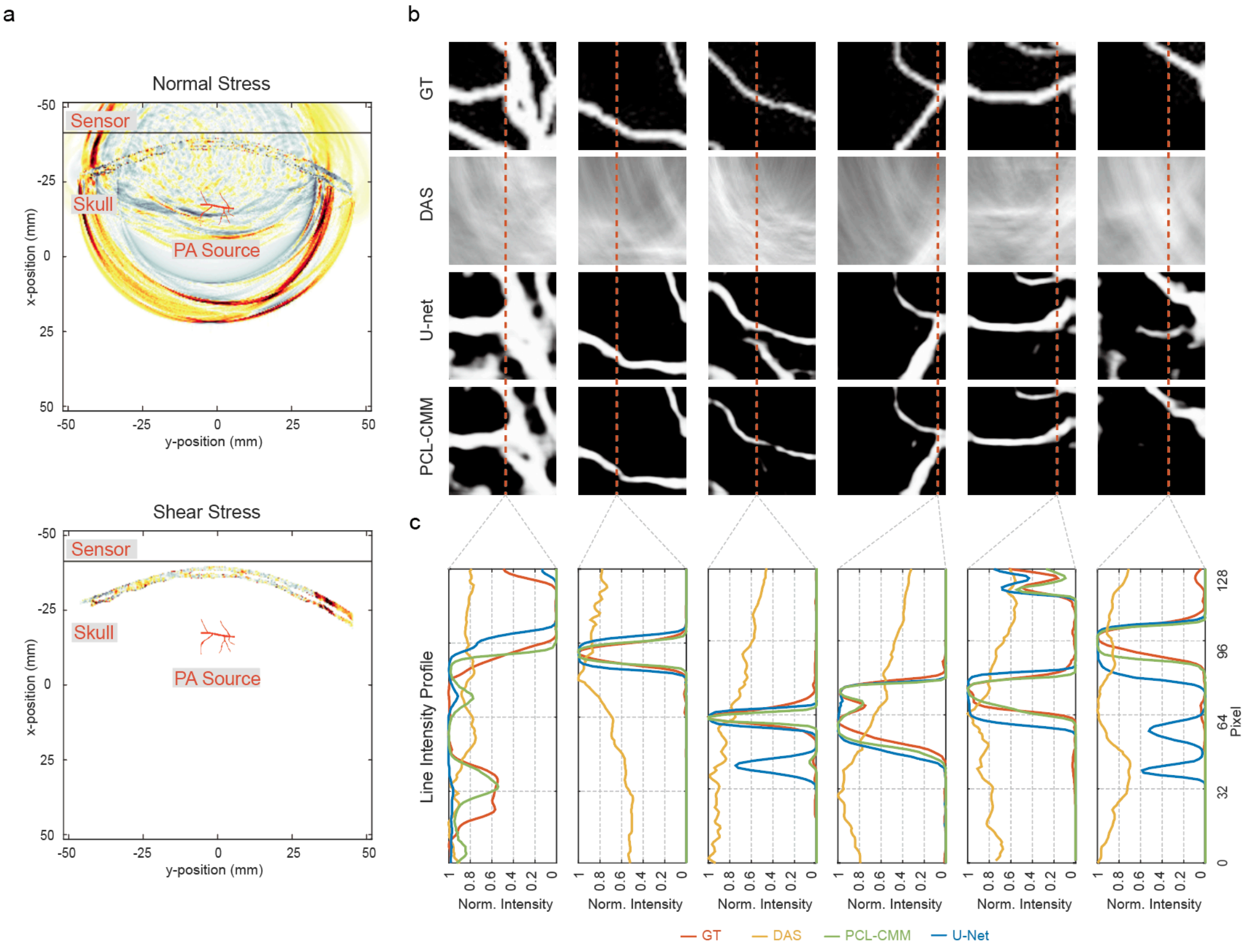

**Fig. 3. Numerical validation and performance comparison of the PCL-CMM framework.** a) Simulation setup illustrating the PA source, skull layer, sensor array, and the propagating wavefield. b) Comparison of reconstruction results among GT, DAS, U-Net, and PCL-CMM. Red dashed lines indicate the locations of intensity profiles. c) Axial intensity profiles corresponding to the red dashed lines in panel b.

**Experimental Validation and Performance Assessment**

To validate the effectiveness of the PCL-CMM framework in a real physical environment, we established an ex vivo experimental platform using an human skull sample (measurement location on the temporal bone, thickness 3.87 mm), an ultrasound array, and a vascular phantom (Fig. 4a). Detailed photographs of the experimental setup and the CT scans of the skull used for validation are provided in Supplementary Fig. 2. The comparison of reconstruction results (Fig. 4b) intuitively demonstrates the performance differences among the algorithms in handling complex transcranial distortions. Although the DAS algorithm can identify the main stem vessels, the reconstruction is plagued by heavy background noise due to severe phase aberrations induced by the skull, resulting in the fine vascular structures being almost completely masked. While the data-driven U-Net significantly improves reconstruction clarity and restores the basic morphology of the vessels, it still retains non-physical artifacts when dealing with vascular intersections and fine branches. In contrast, the reconstruction results of PCL-CMM exhibit the highest consistency with the GT, effectively suppressing background clutter while clearly presenting complex vascular networks.

By extracting line intensity profiles from the reconstructed images, the resolution capabilities of the algorithms can be more objectively evaluated (Fig. 4c). The profile curves for DAS show large-scale broadening and extremely low amplitudes, reflecting severe energy defocusing and localization failure. While the U-Net curves capture the peaks of the primary vessels, they exhibit a noticeable resolution bottleneck when distinguishing densely distributed parallel vessels. Objective evaluation shows that the intensity curves of PCL-CMM achieve high alignment with the GT; despite some minor residual errors in the smallest amplitude details, its ability to correct wavefront distortions is significantly superior to the comparative methods, enabling precise distinction of vascular branches with very small spacing.

The quantitative evaluation metrics listed in Table 1 further support the visual analysis. At the temporal bone testing position with a thickness of 3.87 mm, PCL-CMM achieved the lowest MAE (0.077) and RMSE (0.243). Its SSIM (0.701) and PSNR (12.516 dB) reached optimal levels, and it obtained a 99% win rate in direct comparison with U-Net.

To further evaluate the generalization ability of the model, we shifted the imaging position by approximately 5 mm for a "Cross Position" generalization test; the corresponding skull CT structure for this test area is detailed in Supplementary Fig. 2f. Notably, when facing this unseen anatomical location, the performance gap between PCL-CMM and the standard U-Net widened compared to the training position: the SSIM lead of PCL-CMM over U-Net expanded from 0.060 at the training position to 0.071, and the PSNR gain significantly increased from 0.228 dB to 0.446 dB. In this highly challenging scenario, the metrics for U-Net exhibited significant degradation, whereas PCL-CMM maintained high reconstruction accuracy and achieved a 100% overwhelming win rate, leveraging its mastery of universal wave physics laws. This result strongly demonstrates that the integration of physical priors allows the model to move beyond over-reliance on specific anatomical data, showing stronger generalization resilience compared to purely data-driven models when faced with anatomical structural changes.

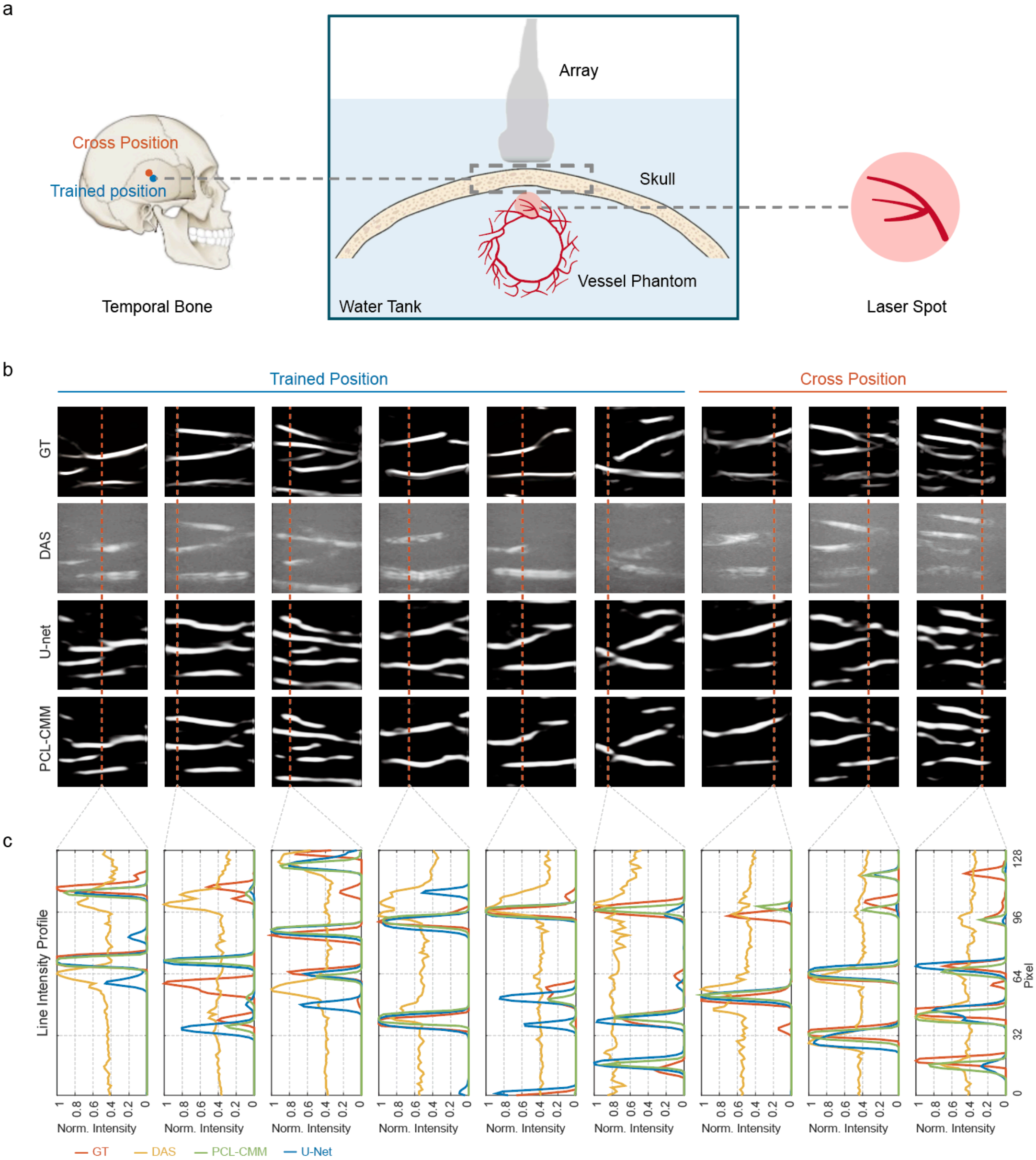

**Fig. 4. Experimental validation and performance comparison of the PCL-CMM framework.** (a) Schematic of the experimental setup showing the ultrasound array, human skull, and vessel phantom in a water tank with designated trained (blue) and cross (red) positions on the temporal bone; (b) visual comparison of reconstruction results for Ground Truth (GT), DAS, U-Net, and PCL-CMM across multiple vascular phantom samples; and (c) corresponding normalized line intensity profiles extracted from the dashed lines in (b) to evaluate reconstruction fidelity and resolution.

**Table 1. Quantitative metrics for simulation and experiment.**

| | Simulation | | | Experiment | | | Experiment (Cross Position) | | |
|---|---|---|---|---|---|---|---|---|---|
| Metrics | DAS | U-Net | PCL-CMM | DAS | U-Net | PCL-CMM | DAS | U-Net | PCL-CMM |
| MAE | 0.525 | 0.110 | 0.082 | 0.300 | 0.087 | 0.077 | 0.351 | 0.101 | 0.086 |
| MSE | 0.325 | 0.084 | 0.060 | 0.102 | 0.062 | 0.059 | 0.131 | 0.073 | 0.067 |
| RMSE | 0.569 | 0.290 | 0.245 | 0.320 | 0.250 | 0.243 | 0.361 | 0.269 | 0.256 |
| SSIM | 0.092 | 0.693 | 0.756 | 0.006 | 0.641 | 0.701 | 0.007 | 0.578 | 0.649 |
| PSNR | 4.765 | 14.376 | 15.648 | 9.925 | 12.288 | 12.516 | 8.847 | 11.470 | 11.916 |
| Win Rate | / | 9% | 91% | / | 1% | 99% | / | 0% | 100% |

## Discussion

The integration of physical constraints via the PCL-CMM framework significantly enhances the fidelity of transcranial photoacoustic reconstruction compared to conventional methodologies. While the DAS algorithm is fundamentally constrained by its inability to compensate for severe phase aberrations, and the standard U-Net relies on a purely data-driven black-box mapping that often yields non-physical artifacts, PCL-CMM ensures adherence to the governing wave physics through a physics-consistency loss. By incorporating the stiffness matrix prior derived from the three-layer skull geometry, the network effectively suppresses heavy background noise and recovers intricate vascular branches that are otherwise masked by bone-induced scattering. This objective alignment with wave propagation laws allows for a more reliable correction of wavefront distortions compared to unconstrained neural networks.

The generalization performance observed at the cross position further underscores the limitations of purely data-driven models, which exhibit a high degree of dependence on the specific anatomical structures encountered during training. Notably, some quantitative metrics obtained from the experimental data are higher than those from the simulations. This difference is mainly attributed to the structural complexity of the vascular patterns: compared with the highly intricate vessel networks from the DRIVE dataset used in simulations, the experimental vessel phantoms were simplified and dominated by transverse vessels because the use of a linear array transducer introduces reduced sensitivity to longitudinal vessels. Future work could involve spherical or hemispherical arrays to achieve omnidirectional acoustic detection and more complete vascular mapping. Future advancements could involve the implementation of spherical or hemispherical arrays[28] to achieve omnidirectional acoustic detection and more complete vascular mapping.

Despite the promising results, this study is subject to certain limitations that warrant further investigation. Our experimental validation and generalization tests were primarily conducted at the temporal bone, where the skull thickness (3.87 mm) provides a relatively favorable acoustic window. However, extending the method to thicker anatomical regions will require overcoming a more formidable acoustic skull barrier through coordinated optimization of both optical excitation and acoustic detection. Specifically, the light fluence distribution in the imaging region should be optimized within safety limits, while high-sensitivity, multi-element detector arrays coaxially aligned with laser illumination should be developed to ensure sufficient transcranial signal quality. These hardware improvements are essential for preventing the detected signals from being dominated by noise and for enabling the proposed physics-constrained algorithm to exert its full advantage. A significant hurdle for clinical adoption is the individualized variance in skull properties.

A significant hurdle for clinical adoption is the individualized variance in skull properties. As demonstrated in Fig. 2 and Supplementary Fig. 3, variations in the middle diploe layer porosity and the thicknesses of the constituent layers introduce a distinct drift in the transmittance characteristic spectra. These parameter-induced shifts in resonance peaks and transmission windows represent the complex, subject-specific acoustic environments that often cause purely data-driven models to fail when encountering unseen anatomical data. Nevertheless, a critical observation from our spectral analysis is that despite these shifts, the transmittance characteristics still maintain a fundamental physical consistency governed by the laws of wave propagation in multi-layered media. This inherent alignment with wave physics is the primary reason why the PCL-CMM framework achieves superior performance at the cross position. By incorporating the stiffness-matrix-based prior, the network successfully leverages this underlying consistency to accommodate characteristic drifts rather than being confined to a narrow, static anatomical mapping. In future work, to address individualized differences, the problem of extreme inter-subject variability in skull morphology could be solved starting from the implementation of subject-specific adaptive modules that dynamically calibrate stiffness matrix parameters based on localized structural data or real-time acoustic feedback.

Finally, the applicability of the PCL-CMM framework extends beyond transcranial medical imaging to a broader range of complex, multi-layered media. The transmittance coefficient analysis in Fig. 2 demonstrates that the framework is finely tuned to the physics of high impedance mismatch and porous structures. Such characteristics are not unique to the human skull but are prevalent in various domains, including biological tissues, geological formations, and advanced composite materials. In any scenario where wave propagation is modulated by complex internal geometries and significant acoustic impedance gradients, the integration of stiffness-matrix-based physical priors offers a powerful tool for solving challenging inverse problems, paving the way for high-resolution imaging and detection in diverse scientific and industrial fields.

## Methods

### Theoretical model for wave propagation in complex multilayered media

A two-dimensional layered geometry is considered, consisting of three representative types of media: fluids, isotropic elastic solids and isotropic poroelastic solids. Each layer is laterally homogeneous and bounded by planar interfaces. Fluid layers are governed by the acoustic wave equation[29], elastic solid layers by the Navier–Helmholtz equation[30], and poroelastic layers by Biot's poroelastic theory[9], so that the three layer types support longitudinal waves (in fluids), longitudinal and transverse waves (in elastic solids), and fast/slow longitudinal waves together with transverse waves (in poroelastic solids), respectively. The complete set of governing equations and constitutive relations is provided in Supplementary Note 1.

In the frequency domain and for a given horizontal wavenumber, the displacement and stress fields within each layer are expressed as a superposition of local wave modes. From this representation, a layer characteristic matrix $\boldsymbol{B}_i$ is constructed. By eliminating the modal amplitudes, a layer stiffness matrix $\boldsymbol{K}_i$ is derived, which relates displacement-type variables to traction-type variables at the top and bottom boundaries of the layer. This stiffness matrix approach[26,31] offers enhanced numerical stability compared to conventional transfer-matrix methods, particularly at high frequencies and for thick layers (Supplementary Note 2).

Adjacent layers are coupled through interface continuity conditions. At fluid–solid interfaces, the normal components of displacement and traction are continuous, while the transverse components of traction on the fluid side vanishes. At elastic–poroelastic interfaces, continuity of the solid displacement and total stress is imposed. When the pores are sealed, fluid motion across the interface is suppressed, and the effect of the pore fluid enters only through the total stress response. These continuity conditions are formulated using boundary matrices $\boldsymbol{V}_i$ and integrated with a Rokhlin–Wang–type recursive algorithm to assemble the individual layer stiffness matrices into a global stiffness matrix $K_{G,N}$ for an $N$-layer stack, which relates displacements and tractions at the external top and bottom external boundaries.

The combination of the global compliance matrix derived from $K_{G,N}^{-1}$ with the boundary conditions for an incident plane wave in the top fluid medium and a transmitted wave in the bottom fluid medium yields complex reflection and transmission coefficients as functions of frequency and incidence angle. By sweeping over a predefined frequency–angle grid, the frequency- and angular–dependent transmission and reflectivity response of the multilayered medium is generated. This response is reparameterized as an effective acoustic stiffness tensor-termed the stiffness prior-which compactly encodes the wave-modulation characteristics of the multilayered structure. The stiffness prior serves as a physics constrained prior in the subsequent learning framework.

**Physics-constrained learning framework via k-space spectral modulation**

To decode acoustic signals modulated by complex multilayered structures and solve the associated inverse problem, we proposed the Physics-Constrained Learning framework for Complex Multilayered Media (PCL-CMM). The core of this framework is a deep non-linear mapping operator $\mathcal{G}$ based on a modified U-Net architecture (detailed in Supplementary Fig. 4), designed to recover measured complex wave fields directly into the target object domain. The architecture utilizes a hierarchical encoder to extract multi-scale spatial features and skip connections to transfer high-resolution information to the decoder, ensuring precise recovery of object boundaries and fine structural details despite severe signal aberrations caused by strong medium heterogeneity.

Distinct from conventional black-box models, our framework establishes a physical consistency constraint through a forward operator $\mathcal{P}$, as illustrated in Fig. 2b. During training iterations, the predicted object $\hat{y}$ generated by the network is first converted into the lateral spectrum $p_0(k_x, z)$ via the Fast Fourier Transform (FFT). This spectrum is then subjected to a spectral modulation operator using the stiffness prior, which is dynamically generated. By performing complex-domain tensor multiplication, this operator compactly simulates the energy redistribution of acoustic waves as they propagate through multilayered media, such as the human skull. This process maps the initial pressure field to a structure-modulated spectrum $p_{skull}(k_x, z)$, which is subsequently transformed back into the reconstructed wave field $p_{skull}(x, z)$, using the Inverse Fast Fourier Transform (IFFT).

To enhance optimization stability and filter experimental noise, a Gaussian blurring operator $\mathcal{B}$ is applied to extract low-frequency components during the calculation of the physical consistency error. The physics-constrained regularization term is defined as follows:

$$L_{phys} = \left\| \mathcal{B}\big(\mathcal{P}(\hat{y})\big) - \mathcal{B}(x) \right\|_1$$

The total objective function is a hybrid formulation that combines pixel-level supervision—weighted Binary Cross-Entropy $L_{BCE}$ and Dice loss $L_{Dice}$ —to balance data-driven accuracy with physical fidelity:

$$L_{total} = L_{BCE} + \delta L_{Dice} + \eta(t) L_{phys}$$

The model is optimized using a progressive training curriculum: during the initial $T_{warm}$ epochs, the regularization weight $\eta(t)$ is set to zero to allow the network to establish basic feature mappings. Subsequently, $\eta(t)$ is linearly ramped to its maximum value, $\eta_{max}$ Throughout this process, the spectral modulation operator serves as a structural regularizer; by maintaining a differentiable computational graph, physical discrepancies are backpropagated as gradients to the neural network. This guides the model to effectively compensate for acoustic aberrations introduced by interlayer discontinuities and porosity, achieving high-fidelity reconstruction in complex computational imaging tasks.

**Numerical Simulation Setup**

To evaluate the performance of the PCL-CMM framework, we conducted 2D numerical simulations using the k-Wave toolbox[32]. Vascular structures from the DRIVE dataset were utilized as the initial PA sources, while the complex multilayered geometry was derived from CT images of the human skull. The simulation domain was discretized into a 1024 × 1024 grid with a spatial resolution of 100 µm per grid point.

The forward propagation of transcranial PA signals was modeled using elastic wave simulation, accounting for the conversion between longitudinal and shear waves within the bone layers. To maintain numerical stability and accurately capture the high-frequency acoustic field, a time step of 6 ns was employed. The specific acoustic and elastic properties of the medium were assigned to the fluid, elastic solid, and porous solid layers according to the values detailed in Supplementary Table 1[33–36]. The relative spatial arrangement of the PA sources, the three-layer skull structure, and the sensor array follows the configuration shown in Fig. 3a.

After capturing the transcranial vascular PA signals at the sensor array, the classical DAS algorithm was employed to reconstruct the initial pressure field. This initial reconstruction served as the input data for the subsequent neural network.

**Experimental Setup and Specimen Preparation**

In this study, experimental data acquisition was conducted using a self-developed PAI system[37]. The laser excitation was provided by a pulsed laser system (Phocus Mobile, OPOTEK, Carlsbad, CA) operating at a wavelength of 700 nm. The system delivered pulses at a repetition rate of 10 Hz, with a single pulse duration of 2–5 ns, ensuring efficient photoacoustic generation under thermal and stress confinement conditions.

Acoustic signals were detected using a high-density 128-element linear ultrasound transducer array. The probe features a center frequency of 10 MHz with a functional frequency range of 6–15 MHz and a detection bandwidth exceeding 90%. The geometric parameters of each array element were precisely defined, with a length of 3.5 mm, a width of 0.2 mm, and an inter-element spacing of 0.2 mm, facilitating high-resolution spatial sampling and signal fidelity.

Prior to acoustic testing and CT scanning, the human temporal bone specimen was subjected to a 24-hour degassing process to eliminate entrapped air within the porous bone matrix, thereby approximating its in vivo structural state and ensuring stable acoustic coupling. All procedures involving human skull samples were conducted in accordance with ethical standards and were formally approved by the institutional review board under Tongji University Ethical Approval No. tjdxsr2024079. The specimen was scanned using a clinical CT system (uCT 960+, United Imaging Healthcare, Shanghai, China) at a spatial resolution of $0.1846 \times 0.1846 \times 0.5\ mm^3$. The resulting CT data were used to obtain the structural information as input to physics-consistency loss function.

**Network Training and Computational Implementation**

The training and evaluation of the proposed framework were conducted using a simulation dataset comprising 500 images and an experimental dataset of 510 images. For both data sources, the total samples were partitioned into training, validation, and testing sets following a fixed ratio of 7:1:2 to ensure robust model selection and reliable final evaluation. To maintain a rigorous and unbiased performance comparison, the data-driven U-Net and the physics-constrained PCL-CMM framework were trained and tested using identical datasets and splits. Optimization was carried out using the adam optimizer, with the total number of training iterations ranging from 40 to 60 epochs to achieve convergence. The Win Rate metric, utilized for the comparative assessment between the algorithms, was calculated based on the SSIM of each reconstructed sample within the test sets. All numerical calculations and model training procedures were performed on an NVIDIA RTX 6000 GPU workstation, utilizing the CUDA 12.2 toolkit and the PyTorch 2.2.2 framework.

## Acknowledgements

This project was supported by the National Natural Science Foundation of China (12034015, 62088101), Program of Shanghai Academic Research Leader (21XD1403600), Shanghai Municipal Science and Technology Major Project (2021SHZDZX0100).

## Supplementary Note 1: Theoretical Framework for Wave Propagation in Cranial Media

1.1 Layer characteristic matrix in Isotropic Fluid Media (Soft Tissue)

Soft tissues, such as the brain and scalp, are modeled as isotropic fluids where the shear modulus vanishes ($\mu = 0$) and the bulk modulus is given by first Lamé parameter： $K_f = \lambda$. The dynamics of the fluid displacement vector $\boldsymbol{u}$ are governed by the wave equation[29]:

$$\rho \frac{\partial^2 \boldsymbol{u}}{\partial t^2} = K_f \nabla(\nabla \cdot \boldsymbol{u}) \tag{S1}$$

where $\rho$ is the density, $t$ denotes time, $\nabla$ is the gradient operator, $\nabla \cdot$ is the divergence operator, and boldface symbols denote vectors and matrices throughout this work.

Analogous to the solid case, the motion is described by a scalar potential $\phi$, which satisfies the wave equation:

$$\frac{\partial^2 \phi}{\partial t^2} - \frac{1}{c_L^2} \nabla^2 \phi = 0, \qquad c_L = \sqrt{\frac{\lambda}{\rho}}, \tag{S2}$$

where $c_L$ denotes the longitudinal wave speed. In fluid media, the acoustic pressure $P_f$ is directly proportional to the volumetric strain (divergence of the displacement field):

$$P_f = -K_f(\nabla \cdot \boldsymbol{u}). \tag{S3}$$

In the context of the multilayered model, the total displacement field $\boldsymbol{u}_i$ within the $i$-th layer (considering components in the $xz$ plane) is constructed as a superposition of all propagating wave modes:

$$\boldsymbol{u}_i(x,z) = \sum_m a_i^m \boldsymbol{p}_i^m e^{j(\boldsymbol{k}_i^m \cdot \boldsymbol{r} - \omega t)}, \tag{S4}$$

where $a_i^m$ denotes the amplitude of the $m$-th wave mode in the $i$-th layer, $\boldsymbol{p}_i^m$ is the displacement polarization vector, $\boldsymbol{k}_i^m$ is the wave vector, $\boldsymbol{r}$ is the position vector, and $\omega$ is the angular frequency. The wavenumber $k_i^m$ corresponds to the specific phase velocity $c_m$:

$$k_i^m = \frac{\omega}{c_m} \tag{S5}$$

Based on Snell's law, the horizontal wavenumber $k_x$ remains invariant across all layers to satisfy phase matching at the interfaces:

$$k_{x_i}^m = k_0 \sin\theta_0 = k_x, \tag{S6}$$

where $k_0$ and $\theta_0$ are the wavenumber and incident angle in the coupling fluid, respectively. Assuming the system is invariant in the $z$-direction, the vertical wavenumber for the $m$-th mode in the $i$-th layer is derived as:

$$k_{z_i}^{m^+} = -k_{z_i}^{m^-} = \sqrt{(k_i^m)^2 - (k_x)^2}. \tag{S7}$$

Let $\boldsymbol{A}_i$ denote the column vector of wave amplitudes in the $i$-th layer. We introduce the layer characteristic matrix $\boldsymbol{B}_i$, which is determined solely by the layer's material properties and wave vector orientation, and the diagonal matrix $\boldsymbol{E}_i(z)$, which governs the spatial propagation and directionality along the $z$-axis. By factorizing the local wave amplitudes using these matrices, the state vector $\boldsymbol{U}_i$ is explicitly formulated as:

$$\boldsymbol{U}_i(x,z) = [\boldsymbol{B}_i][\boldsymbol{E}_i(z)]\boldsymbol{A}_i e^{j(k_x x-\omega t)} \quad z_i \leq z \leq z_{i+1}. \tag{S8}$$

Reflecting the continuity of normal displacement and acoustic pressure at fluid interfaces, the physical state vector $\boldsymbol{U}_i \in \mathbb{C}^{2\times 1}$ for the $i$-th fluid layer (see Supplementary Fig. 1a), along with its constituent characteristic matrix $\boldsymbol{B}_i \in \mathbb{C}^{2\times 2}$ and diagonal propagation matrix $\boldsymbol{E}_i \in \mathbb{C}^{2\times 2}$, are explicitly formulated as:

$$\boldsymbol{U}_i(x,z) = \begin{bmatrix} u_{z_i}(x,z) \\ -P_{f_i}(x,z) \end{bmatrix}, \tag{S9}$$

$$[\boldsymbol{B}_i] = \begin{bmatrix} p_{z_i}^{L+} & p_{z_i}^{L-} \\ jK_{f_i}\left(p_{z_i}^{L+}k_{z_i}^{L+} + p_{x_i}^{L+}k_x\right) & jK_{f_i}\left(p_{z_i}^{L-}k_{z_i}^{L-} + p_{x_i}^{L-}k_x\right) \end{bmatrix}, \tag{S10}$$

$$[\boldsymbol{E}_i(z)] = \begin{bmatrix} e^{jk_{z_i}^{L+}(z-z_i)} & 0 \\ 0 & e^{jk_{z_i}^{L-}(z-z_{i+1})} \end{bmatrix}. \tag{S11}$$

1.2 Layer characteristic matrix in Isotropic Elastic Solids (Cortical Bone)

The cortical bone is modeled as an isotropic, linearly elastic solid characterized by its mass density $\rho$ and Lamé parameters $\lambda$ and $\mu$. Governed by Newton's second law of motion and the linear stress-strain constitutive relation, the displacement vector $\boldsymbol{u}$ satisfies the Navier-Helmholtz equation in the frequency domain[30]:

$$\rho\frac{\partial^2 \boldsymbol{u}}{\partial t^2} = (\lambda+\mu)\nabla(\nabla\cdot\boldsymbol{u}) + \mu\nabla^2\boldsymbol{u}, \tag{S12}$$

where $\nabla^2$ is the Laplacian operator. To decouple the longitudinal (P) and transverse (S) waves, the Helmholtz decomposition can be applied, expressing the displacement field as the sum of the gradient of a scalar potential $\phi$ and the curl of a vector potential $\boldsymbol{\psi}$:

$$\boldsymbol{u} = \nabla\phi + \nabla\times\boldsymbol{\psi}, \tag{S13}$$

which is subjected to the gauge condition:

$$\nabla\cdot\boldsymbol{\psi} = 0. \tag{S14}$$

Substituting Eq. (2) into Eq. (1) yields two decoupled scalar wave equations governing $\phi$ and $\boldsymbol{\psi}$:

$$\begin{cases} \dfrac{\partial^2\phi}{\partial t^2} - \dfrac{1}{c_L^2}\nabla^2\phi = 0, \\ \dfrac{\partial^2\boldsymbol{\psi}}{\partial t^2} - \dfrac{1}{c_T^2}\nabla^2\boldsymbol{\psi} = 0. \end{cases} \tag{S15}$$

Here, $c_L$ and $c_T$ represent the propagation speeds of the longitudinal and transverse waves, respectively, defined as:

$$c_L = \sqrt{\frac{\lambda + 2\mu}{\rho}}, \qquad c_T = \sqrt{\frac{\mu}{\rho}}. \tag{S16}$$

For an isotropic elastic solid layer, four wave modes exist: downward/upward longitudinal waves and downward/upward transverse waves $(m^+ = L^+, T^+;\ m^- = L^-, T^-)$ (see Supplementary Fig. 1b). The normalized polarization vectors $\boldsymbol{p}_i^m$ for these modes are expressed as:

$$\begin{gathered} \boldsymbol{p}_i^{L^+} = [\sin\theta_i^L, 0, \cos\theta_i^L]^{\mathrm{T}};\ \boldsymbol{p}_i^{L^-} = [\sin\theta_i^L, 0, -\cos\theta_i^L]^{\mathrm{T}}; \\ \boldsymbol{p}_i^{T^+} = [-\cos\theta_i^T, 0, \sin\theta_i^T]^{\mathrm{T}};\ \boldsymbol{p}_i^{T^-} = [\cos\theta_i^T, 0, \sin\theta_i^T]^{\mathrm{T}}. \end{gathered} \tag{S17}$$

Reflecting the continuity of the displacement components ($x$ and $z$) and the stress tensor components ($xz$ and $zz$) at the solid interfaces, the physical state vector $\boldsymbol{U}_i \in \mathbb{C}^{4\times1}$, along with its constituent characteristic matrix $\boldsymbol{B}_i \in \mathbb{C}^{4\times4}$ and diagonal propagation matrix $\boldsymbol{E_i} \in \mathbb{C}^{4\times4}$, are explicitly formulated as:

$$\boldsymbol{U}_i(x,z) = \begin{bmatrix} u_{x_i}(x,z) \\ u_{z_i}(x,z) \\ \sigma_{xz_i}(x,z) \\ \sigma_{zz_i}(x,z) \end{bmatrix}, \tag{S18}$$

$$[\boldsymbol{B}_i]_{4\times4} = \begin{bmatrix} p_{x_i}^m \\ p_{z_i}^m \\ j\mu_i\left(p_{x_i}^m k_{z_i}^m + p_{z_i}^m k_x\right) \\ j\left[(\lambda_i + 2\mu_i) p_{z_i}^m k_{z_i}^m + \lambda_i p_{x_i}^m k_x\right] \end{bmatrix}, \tag{S19}$$

$$\mathrm{diag}\{[\boldsymbol{E}_i(z)]\} = \left[e^{jk_{z_i}^{L^+}(z-z_i)}\ e^{jk_{z_i}^{T^+}(z-z_i)}\ e^{jk_{z_i}^{L^-}(z-z_{i+1})}\ e^{jk_{z_i}^{T^-}(z-z_{i+1})}\right]. \tag{S20}$$

1.3 Layer characteristic matrix in Isotropic Poroelastic Media (Diploe Bone)

The diploe layer is modeled as a fluid-saturated isotropic porous medium governed by Biot's theory[9]. The coupled equations of motion describing the solid skeletal displacement $\boldsymbol{u}$ and the relative fluid displacement $\boldsymbol{w}$ are given by:

$$\begin{gathered} \nabla\cdot\boldsymbol{\sigma} = H_c\nabla(\nabla\cdot\boldsymbol{u}) - \mu\nabla\times(\nabla\times\boldsymbol{u}) + D\nabla(\nabla\cdot\boldsymbol{w}) = \rho\frac{\partial^2\boldsymbol{u}}{\partial t^2} + \rho_f\frac{\partial^2\boldsymbol{w}}{\partial t^2}, \\ -\nabla P_f = D\nabla(\nabla\cdot\boldsymbol{u}) + M\nabla(\nabla\cdot\boldsymbol{w}) = \rho_f\frac{\partial^2\boldsymbol{u}}{\partial t^2} + g\frac{\partial^2\boldsymbol{w}}{\partial t^2} + b\frac{\partial\boldsymbol{w}}{\partial t}, \end{gathered} \tag{S21}$$

where $\boldsymbol{\sigma}$ is the total stress tensor and $P_f$ is the pore fluid pressure. The physical properties are defined by the porosity $\phi$, fluid density $\rho_f$, bulk modulus $K_f$, solid grain density $\rho_s$, moduli $K_s$, $\mu_s$, the skeletal frame properties $K_m$, $\mu_m$ (bulk and shear moduli), permeability $\kappa_m$. The Biot's effective parameters ($H_c$, $M$, $D$, etc.) appearing in the constitutive relations are derived as follows:

$$H_c = \lambda_c + 2\mu, \tag{S22}$$

$$\lambda_c = K_c - \frac{2}{3}\mu,$$

$$K_c = K_m + \alpha^2 M,$$

$$\alpha = 1 - \frac{K_m}{K_s},$$

$$M = \left(\frac{\alpha - \phi}{K_s} + \frac{\phi}{K_f}\right)^{-1},$$

$$D = \alpha M,$$

$$\rho = (1 - \phi)\rho_s + \phi\rho_f,$$

$$g = \frac{S\rho_f}{\phi},$$

$$b = \frac{\eta_f}{\kappa_m},$$

$$S = \frac{1}{2}\left(1 + \frac{1}{\phi}\right).$$

In such media, wave propagation consists of three pairs of wave modes: fast longitudinal ($L_1^+, L_1^-,$), slow longitudinal ($L_2^+, L_2^-$), and transverse ($T^+, T^-$) waves (see Supplementary Fig. 1c).

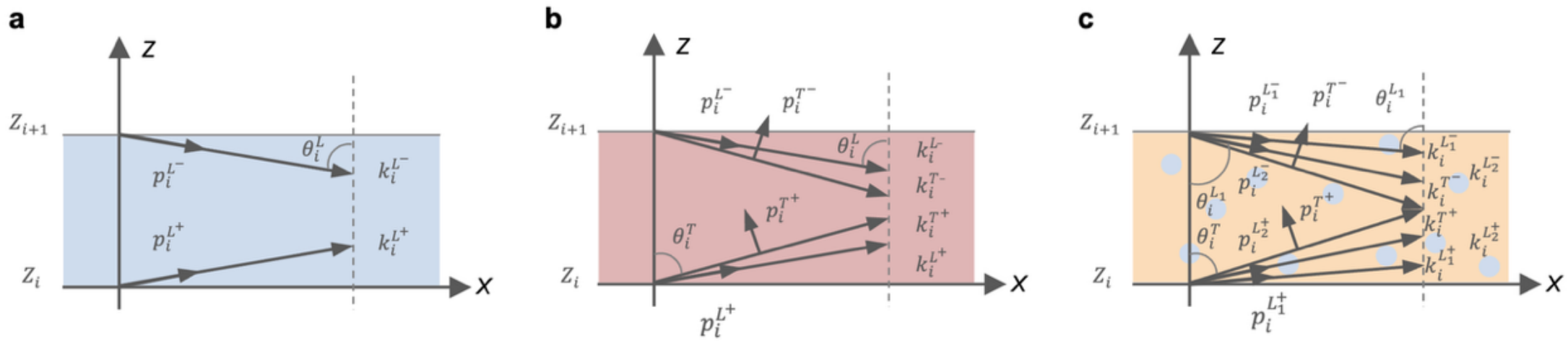

**Supplementary Fig. 1.** Schematic of wave propagation modes in different media layers. a) Wave propagation in an isotropic fluid layer (representing soft tissue), which supports only downward (+) and upward (−) propagating longitudinal ($L$) waves. b) Propagation in an isotropic elastic solid layer (representing cortical bone), involving both longitudinal ($L$) and transverse ($T$) wave modes. c) Propagation in an isotropic poroelastic layer (representing diploe) governed by Biot's theory. This medium supports three pairs of wave modes: fast longitudinal ($L_1$), slow longitudinal ($L_2$), and transverse ($T$) waves. In all panels, vectors $\boldsymbol{p}_i^m$ and $\boldsymbol{k}_i^m$ denote the displacement polarization vector and wave vector, respectively, for mode $m$ in the $i$-th layer. The $\theta_i^m$ represents the propagation angle relative to the $z$-axis.

For convenience in solving the characteristic equation, we introduce three auxiliary quantities $G, F$ and $Q$, which are employed to simplify the algebraic form of the governing equation. By applying the Helmholtz decomposition, the wavenumbers for the longitudinal modes are determined by the characteristic equation:

$$G(k^L)^4 + F(k^L)^2 + Q = 0, \tag{S23}$$

where the auxiliary quantities are given by:

$$\begin{gathered} G = H_c M - D^2, \\ F = \omega^2\left(-H_c g - M\rho + 2D\rho_f\right) + j\omega H_c b, \\ Q = \omega^4\left(\rho g - \rho_f^2\right) - j\omega^3 \rho b. \end{gathered} \tag{S24}$$

Solving Eq. (23) yields two distinct roots corresponding to the fast and slow longitudinal wavenumbers:

$$k^{L_{12}} = \sqrt{\frac{-F \pm \sqrt{F^2 - 4GQ}}{2G}}. \tag{S25}$$

For the longitudinal modes, the amplitudes of the potentials of the solid and the relative fluid displacements are denoted by $X^L$ and $Y^L$, respectively. The relationship between this two is expressed as:

$$Y^L = \gamma^L X^L,\ \gamma^L = \frac{-H_c(k^L)^2 + \rho\omega^2}{D(k^L)^2 - \rho_f \omega^2}. \tag{S26}$$

Similarly, for the transverse modes, the wavenumber $k^T$ and the fluid-solid amplitude ratio $\gamma^T$ are derived as:

$$k^T = \sqrt{\frac{\omega^2}{\mu}\left(\rho - \frac{\rho_f^2}{g - \frac{jb}{\omega}}\right)}, \tag{S27}$$

$$Y^T = \gamma^T X^T,\ \gamma^T = \frac{\mu(k^T)^2 + \rho\omega^2}{\rho_f \omega^2}. \tag{S28}$$

The boundary conditions at the poroelastic interfaces require the continuity of solid displacements ($u_x$, $u_z$), normal fluid displacement ($w_z$), total stresses ($\sigma_{xz}$, $\sigma_{zz}$), and pore pressure ($P_f$). Accordingly, the state vector $\boldsymbol{U}_i \in \mathbb{C}^{6\times 1}$ for the $i$-th isotropic poroelastic layer is defined as:

$$\boldsymbol{U}_i(x,z) = [u_{x_i}(x,z) \quad u_{z_i}(x,z) \quad w_{z_i}(x,z) \quad \sigma_{xz_i}(x,z) \quad \sigma_{zz_i}(x,z) \quad P_{f_i}(x,z)]^{\mathrm{T}}, \tag{S29}$$

$$\begin{gathered} \mathrm{diag}\{[E_i(z)]\} = [e^{jk_{z_i}^{L_1^+}(z-z_i)} \; e^{jk_{z_i}^{L_2^+}(z-z_i)} \; e^{jk_{z_i}^{T^+}(z-z_i)}, \\ e^{jk_{z_i}^{L_1^-}(z-z_{i+1})} \; e^{jk_{z_i}^{L_2^-}(z-z_{i+1})} \; e^{jk_{z_i}^{T^-}(z-z_{i+1})}], \end{gathered} \tag{S30}$$

$$[\boldsymbol{B}_i] = \begin{bmatrix} p_{x_i}^m \\ p_{z_i}^m \\ \gamma_i^m p_{z_i}^m \\ [\boldsymbol{C}_i^m] \end{bmatrix}, \qquad m = L_1^{\pm}, L_2^{\pm}, T^{\pm}, \tag{S31}$$

$$[\boldsymbol{C}_i^m] = \begin{cases} \begin{bmatrix} j2\mu_i k_i^m p_{x_i}^m p_{z_i}^m \\ k_i^m \left(2\mu_i \left(p_{z_i}^m\right)^2 + \lambda_{c_i} + D_i \gamma_i^m\right) \\ -j(D_i + \gamma_i^m M_i) k_i^m \end{bmatrix}, \text{for } m = L_1^\pm, L_2^\pm, \\ \begin{bmatrix} j\mu_i k_i^m \left(\left(p_{z_i}^m\right)^2 - \left(p_{x_i}^m\right)^2\right) \\ -j2\mu_i k_i^m p_{x_i}^m p_{z_i}^m \\ 0 \end{bmatrix}, \quad \text{for } m = T^\pm. \end{cases} \tag{S32}$$

**Supplementary Note 2: Stiffness Matrix Method and Recursive Solution**

To avoid numerical instability in the transfer matrix method at high frequencies or for thick layers, we use the Stiffness Matrix Method (SMM)[26,31]. This method defines the layer stiffness matrix by linking the displacement and stress vectors at the top and bottom surfaces of a single layer.

For the $i$-th layer, we divide the state vector into two parts: the displacement vector $\boldsymbol{v}_i$ and the stress vector $\boldsymbol{T}_i$. We use $n_i = 1,2,3$ to represent fluid, solid, and porous solid layers, respectively. The layer characteristic matrix is split into four equal sub-matrices:

$$[\boldsymbol{B}_i]_{2n_i\times 2n_i} = \begin{bmatrix} [\boldsymbol{B}_i^{11}]_{n_i\times n_i} & [\boldsymbol{B}_i^{12}]_{n_i\times n_i} \\ [\boldsymbol{B}_i^{21}]_{n_i\times n_i} & [\boldsymbol{B}_i^{22}]_{n_i\times n_i} \end{bmatrix}. \tag{S33}$$

Using these sub-matrices, we can write $\boldsymbol{v}_i$ and $\boldsymbol{T}_i$ as:

$$\begin{aligned} v_i(x,z) &= \begin{bmatrix}[\boldsymbol{B}_i^{11}] & [\boldsymbol{B}_i^{12}]\end{bmatrix}[\boldsymbol{E}_i(z)]\boldsymbol{A}_i e^{j(k_x x)}, \\ \boldsymbol{T}_i(x,z) &= \begin{bmatrix}[\boldsymbol{B}_i^{21}] & [\boldsymbol{B}_i^{22}]\end{bmatrix}[\boldsymbol{E}_i(z)]\boldsymbol{A}_i e^{j(k_x x)}. \end{aligned} \tag{S34}$$

The displacements at the top and bottom surfaces of the $i$-th layer are given by:

$$\begin{bmatrix} \boldsymbol{v}_i(x,z_i) \\ \boldsymbol{v}_i(x,z_{i+1}) \end{bmatrix} = \begin{bmatrix} [\boldsymbol{B}_i^{11}] & [\boldsymbol{B}_i^{12}][\boldsymbol{H}_i^-] \\ [\boldsymbol{B}_i^{11}][\boldsymbol{H}_i^+] & [\boldsymbol{B}_i^{12}] \end{bmatrix} \begin{bmatrix} \boldsymbol{A}_i^+ \\ \boldsymbol{A}_i^- \end{bmatrix} e^{j(k_x x)} = [\boldsymbol{E}_i^v]\boldsymbol{A}_i e^{jk_x x}. \tag{S34}$$

Here, $[\boldsymbol{H}_i^+]$ and $[\boldsymbol{H}_i^-]$ are diagonal matrices: $\text{diag}\{[\boldsymbol{H}_i^+]\} = e^{jk_{z_i}^{m^+} h_i}$ and $\text{diag}\{[\boldsymbol{H}_i^-]\} = e^{-jk_{z_i}^{m^-} h_i}$, where $h_i$ is the layer thickness. $\boldsymbol{A}_i^+$ and $A_i^-$ are the amplitude vectors for waves moving in the $+z$ and $-z$ directions. For a system symmetric about the $z$-axis, we have $[\boldsymbol{H}_i^+] = [H_i^-]$.

Similarly, the stress components are written as:

$$\begin{bmatrix} \boldsymbol{T}_i(x,z_i) \\ \boldsymbol{T}_i(x,z_{i+1}) \end{bmatrix} = \begin{bmatrix} [\boldsymbol{B}_i^{21}] & [\boldsymbol{B}_i^{22}][\boldsymbol{H}_i^-] \\ [\boldsymbol{B}_i^{21}][\boldsymbol{H}_i^+] & [\boldsymbol{B}_i^{22}] \end{bmatrix} \begin{bmatrix} \boldsymbol{A}_i^+ \\ \boldsymbol{A}_i^- \end{bmatrix} e^{j(k_x x)} = [\boldsymbol{E}_i^T]\boldsymbol{A}_i e^{jk_x x}. \tag{S35}$$

Therefore, the relationship between displacements and stresses at the surfaces is:

$$\begin{aligned} \begin{bmatrix} \boldsymbol{T}_i(x,z_i) \\ \boldsymbol{T}_i(x,z_{i+1}) \end{bmatrix} &= [\boldsymbol{E}_i^T][\boldsymbol{E}_i^v]^{-1} \begin{bmatrix} \boldsymbol{v}_i(x,z_i) \\ \boldsymbol{v}_i(x,z_{i+1}) \end{bmatrix} = [\boldsymbol{K}_i] \begin{bmatrix} \boldsymbol{v}_i(x,z_i) \\ \boldsymbol{v}_i(x,z_{i+1}) \end{bmatrix}, \\ \begin{bmatrix} \boldsymbol{v}_i(x,z_i) \\ \boldsymbol{v}_i(x,z_{i+1}) \end{bmatrix} &= [\boldsymbol{E}_i^v][\boldsymbol{E}_i^T]^{-1} \begin{bmatrix} \sigma_i(x,z_i) \\ \boldsymbol{\sigma}_i(x,z_{i+1}) \end{bmatrix} = [\boldsymbol{S}_i] \begin{bmatrix} \boldsymbol{T}_i(x,z_i) \\ \boldsymbol{T}_i(x,z_{i+1}) \end{bmatrix}, \end{aligned} \tag{S36}$$

where:

$$[S_i] = [K_i]^{-1}. \tag{S37}$$

For a fluid layer ($n$ = 1), the displacement and stress are:

$$\begin{aligned} v_i(x,z) &= u_{z_i}(x,z), \\ T_i(x,z) &= -P_{f_i}(x,z). \end{aligned} \tag{S38}$$

The stiffness matrix is:

$$[\boldsymbol{K}_i]_{2\times2} = [\boldsymbol{E}_i^{\sigma}][\boldsymbol{E}_i^{u}]^{-1}$$
$$= \begin{bmatrix} B_i^{21} & B_i^{22} e^{-jk_{z_i}^{L^-} h_i} \\ B_i^{21} e^{jk_{z_i}^{L^+} h_i} & B_i^{22} \end{bmatrix} \begin{bmatrix} B_i^{11} & B_i^{12} e^{-jk_{z_i}^{L^-} h_i} \\ B_i^{11} e^{jk_{z_i}^{L^+} h_i} & B_i^{12} \end{bmatrix}^{-1}. \tag{S39}$$

For an elastic solid layer (n = 2), the components are:

$$v_i(x,z) = \begin{bmatrix} u_{x_i}(x,z) \\ u_{z_i}(x,z) \end{bmatrix},$$
$$T_i(x,z) = \begin{bmatrix} \sigma_{xz_i}(x,z) \\ \sigma_{zz_i}(x,z) \end{bmatrix}. \tag{S40}$$

The stiffness matrix is:

$$[\boldsymbol{K}_i]_{4\times4} = [\boldsymbol{E}_i^{\sigma}][\boldsymbol{E}_i^{u}]^{-1} = \begin{bmatrix} [\boldsymbol{B}_i^{21}]_{2\times2} & [\boldsymbol{B}_i^{22}]_{2\times2}[\boldsymbol{H}_i^{-}]_{2\times2} \\ [\boldsymbol{B}_i^{21}]_{2\times2}[\boldsymbol{H}_i^{+}]_{2\times2} & [\boldsymbol{B}_i^{22}]_{2\times2} \end{bmatrix},$$
$$\begin{bmatrix} [\boldsymbol{B}_i^{11}]_{2\times2} & [\boldsymbol{B}_i^{12}]_{2\times2}[\boldsymbol{H}_i^{-}]_{2\times2} \\ [\boldsymbol{B}_i^{11}]_{2\times2}[\boldsymbol{H}_i^{+}]_{2\times2} & [\boldsymbol{B}_i^{12}]_{2\times2} \end{bmatrix}^{-1} \tag{S41}$$

where the diagonal matrices $[\boldsymbol{H}_i^{+}]$ and $[\boldsymbol{H}_i^{-}]$ are:

$$\begin{aligned} \operatorname{diag}\{[\boldsymbol{H}_i^{+}]\} &= \begin{bmatrix} e^{jk_{z_i}^{L^+} h_i} & e^{jk_{z_i}^{T^+} h_i} \end{bmatrix} \\ \operatorname{diag}\{[\boldsymbol{H}_i^{-}]\} &= \begin{bmatrix} e^{-jk_{z_i}^{L^-} h_i} & e^{-jk_{z_i}^{T^-} h_i} \end{bmatrix} \end{aligned} \tag{S42}$$

For the isotropic poroelastic solid layer (n = 3), the displacement and stress components are defined as:

$$\boldsymbol{v}_i(x,z) = \begin{bmatrix} u_{x_i}(x,z) \\ u_{z_i}(x,z) \\ w_{z_i}(x,z) \end{bmatrix}$$
$$\boldsymbol{T}_i(x,z) = \begin{bmatrix} \sigma_{xz_i}(x,z) \\ \sigma_{zz_i}(x,z) \\ P_{f_i}(x,z) \end{bmatrix} \tag{S43}$$

The stiffness matrix is:

$$[\boldsymbol{K}_i]_{6\times6} = [\boldsymbol{E}_i^{\sigma}][\boldsymbol{E}_i^{u}]^{-1} = \begin{bmatrix} [\boldsymbol{B}_i^{21}]_{3\times3} & [\boldsymbol{B}_i^{22}]_{3\times3}[\boldsymbol{H}_i^{-}]_{3\times3} \\ [\boldsymbol{B}_i^{21}]_{3\times3}[\boldsymbol{H}_i^{+}]_{3\times3} & [\boldsymbol{B}_i^{22}]_{3\times3} \end{bmatrix}$$
$$\begin{bmatrix} [\boldsymbol{B}_i^{11}]_{3\times3} & [\boldsymbol{B}_i^{12}]_{3\times3}[\boldsymbol{H}_i^{-}]_{3\times3} \\ [\boldsymbol{B}_i^{11}]_{3\times3}[\boldsymbol{H}_i^{+}]_{3\times3} & [\boldsymbol{B}_i^{12}]_{3\times3} \end{bmatrix}^{-1} \tag{S44}$$

where the diagonal matrices $[\boldsymbol{H}_i^{+}]$ and $[\boldsymbol{H}_i^{-}]$ are:

$$\begin{aligned} \operatorname{diag}\{[\boldsymbol{H}_i^{+}]\} &= \begin{bmatrix} e^{jk_{z_i}^{L_1^+} h_i} & e^{jk_{z_i}^{L_2^+} h_i} & e^{jk_{z_i}^{T^+} h_i} \end{bmatrix} \\ \operatorname{diag}\{[\boldsymbol{H}_i^{-}]\} &= \begin{bmatrix} e^{-jk_{z_i}^{L_1^-} h_i} & e^{-jk_{z_i}^{L_2^-} h_i} & e^{-jk_{z_i}^{T^-} h_i} \end{bmatrix} \end{aligned} \tag{S45}$$

In the skull model, the interface between the soft tissue and the cortical bone is treated as a fluid–solid boundary, whereas the interface between the cortical bone and the diploe is described as an elastic solid–poroelastic boundary. Based on the governing equations of the individual media described above, the coupling between adjacent layers is enforced through the interfacial boundary conditions. At an interface located at $z = z_i$, the general coupling conditions can be written as

$$\begin{cases} \sigma_{zz}^{(1)} = \sigma_{zz}^{(2)}, \\ \sigma_{xz}^{(1)} = \sigma_{xz}^{(2)}, \\ u_z^{(1)} + w_z^{(1)} = u_z^{(2)} + w_z^{(2)}, \\ u_x^{(1)} = u_x^{(2)}\ (bouded\ interface)\ or\ \sigma_{xz} = 0 (sliding\ interface), \\ P_f^{(1)} = P_f^{(2)}\ (open\ pore\ condition)\ or\ w_z = 0\ (sealed\ pore\ condition). \end{cases} \tag{S46}$$

These conditions enforce the continuity of normal and shear stresses, the compatibility of displacement across the interface, and the hydraulic constraint associated with the pore fluid in the poroelastic medium. In the present model, the bonded interface corresponds to the cortical bone–diploe boundary, where both the normal and tangential displacements of the solid matrix are continuous. The sliding interface corresponds to the soft tissue (or brain tissue)–cortical bone boundary, where the shear stress vanishes at the fluid–solid interface. The open-pore condition applies to interfaces between a porous medium and a fluid, where the pore fluid remains hydraulically connected to the external fluid. The sealed-pore condition applies to interfaces between a porous medium and an impermeable solid layer. To get the response of the N-layer system, we use the Rokhlin-Wang recursive algorithm. Let $\boldsymbol{K}_{G_{i-1}}$ be the global stiffness matrix for the first $i-1$ layers, and $\boldsymbol{K}_i$ be the matrix for the $i$-th layer. The subsystems satisfy:

$$\begin{gathered} \begin{bmatrix} \boldsymbol{T}_1(x, z_1) \\ \boldsymbol{T}_{i-1}(x, z_i) \end{bmatrix} = \left[\boldsymbol{K}_{G_{i-1}}\right]_{(n_1+n_{i-1})\times(n_1+n_{i-1})} \begin{bmatrix} \boldsymbol{v}_1(x, z_1) \\ \boldsymbol{v}_{i-1}(x, z_i) \end{bmatrix} \\ \begin{bmatrix} \boldsymbol{T}_i(x, z_i) \\ \boldsymbol{T}_i(x, z_{i+1}) \end{bmatrix} = [\boldsymbol{K}_i]_{2n_i\times 2n_i} \begin{bmatrix} \boldsymbol{v}_i(x, z_i) \\ \boldsymbol{v}_i(x, z_{i+1}) \end{bmatrix} \end{gathered} \tag{S47}$$

To assemble the full stiffness matrix, we define the relationship between the interface displacement vector $[\boldsymbol{v}_{i-1}(x, z_i), \boldsymbol{v}_i(x, z_i)]^T$ and the external system displacement vector $[\boldsymbol{v}_1(x, z_1), \boldsymbol{v}_i(x, z_{i+1})]^T$, using the matrix $[\boldsymbol{V}_i]$:

$$\begin{aligned} \begin{bmatrix} \boldsymbol{v}_{i-1}(x, z_i) \\ \boldsymbol{v}_i(x, z_i) \end{bmatrix} &= [\boldsymbol{V}_i]_{(n_{i-1}+n_i)\times(n_1+n_i)} \begin{bmatrix} \boldsymbol{v}_1(x, z_1) \\ \boldsymbol{v}_i(x, z_{i+1}) \end{bmatrix} \\ &= \begin{bmatrix} [\boldsymbol{V}_i^{11}]_{n_{i-1}\times n_1} & [\boldsymbol{V}_i^{12}]_{n_{i-1}\times n_i} \\ [\boldsymbol{V}_i^{21}]_{n_i\times n_1} & [\boldsymbol{V}_i^{22}]_{n_i\times n_i} \end{bmatrix} \begin{bmatrix} \boldsymbol{v}_1(x, z_1) \\ \boldsymbol{v}_i(x, z_{i+1}) \end{bmatrix} \end{aligned} \tag{S48}$$

Substituting Eq. (48) into Eq. (47), we derive the updated global stiffness matrix $\boldsymbol{K}_{G_i}$:

$$\left[\boldsymbol{K}_{G_i}\right]_{(n_1+n_i)\times(n_1+n_i)} = \begin{bmatrix} \left[\boldsymbol{K}_{G_{i-1}}^{11}\right] + \left[\boldsymbol{K}_{G_{i-1}}^{12}\right][\boldsymbol{V}_i^{11}] & \left[\boldsymbol{K}_{G_{i-1}}^{12}\right][\boldsymbol{V}_i^{12}] \\ [\boldsymbol{K}_i^{21}][\boldsymbol{V}_i^{21}] & [\boldsymbol{K}_i^{21}][\boldsymbol{V}_i^{22}] + [\boldsymbol{K}_i^{22}] \end{bmatrix} \tag{S49}$$

Boundary Matrix for Solid - Porous Interface

When layer $i-1$ is an isotropic elastic solid and layer $i$ is an isotropic porous solid, based on the boundary conditions in Eq. (46) (zero fluid displacement), we define the reduced sub-matrices $\boldsymbol{K}_i^{11'}$ and $\boldsymbol{K}_i^{12'}$ by removing the row or column corresponding to the fluid. We can then obtain the mapping operator:

$$[\boldsymbol{W}_i]_{2\times(n_1+3)} = \left[\left[\boldsymbol{K}_{G_{i-1}}^{22}\right] - \left[\boldsymbol{K}_i^{11'}\right]\right]_{2\times2}^{-1} \left[-\left[\boldsymbol{K}_{G_{i-1}}^{21}\right] \quad \left[\boldsymbol{K}_i^{12'}\right]\right]_{2\times(n_1+3)} \tag{S50}$$

Based on this, the complete boundary matrix $[\boldsymbol{V}_i]$ is constructed by padding the fluid displacement term with zeros:

$$[\boldsymbol{V}_i]_{5\times(n_1+3)} = \begin{bmatrix} [\boldsymbol{W}_i]_{2\times(n_1+3)} \\ [\boldsymbol{W}_i]_{2\times(n_1+3)} \\ [\boldsymbol{0}]_{1\times(n+3)} \end{bmatrix} \tag{S51}$$

Boundary Matrix for Porous - Solid Interface:

When layer $i-1$ is a porous solid, $\boldsymbol{K}_{G_{i-1}}$ represents the porous medium. Applying the sealed-pore condition, we define the reduced global matrices $\boldsymbol{K}_{G_{i-1}}^{21}{}'$ and $\boldsymbol{K}_{G_{i-1}}^{22}{}'$ by excluding the last row and column. The mapping operator is calculated as:

$$[\boldsymbol{W}_i]_{2\times(n_1+2)} = \left[\left[\boldsymbol{K}_{G_{i-1}}^{22}{}'\right] - [\boldsymbol{K}_i^{11}]\right]_{2\times2}^{-1} \left[-\left[\boldsymbol{K}_{G_{i-1}}^{21}{}'\right] \quad [\boldsymbol{K}_i^{12}]\right]_{2\times(n_1+2)} \tag{S52}$$

The complete boundary matrix is given by:

$$[\boldsymbol{V}_i]_{5\times(n_1+2)} = \begin{bmatrix} [\boldsymbol{W}_i]_{2\times(n_1+2)} \\ [\boldsymbol{0}]_{1\times(n_1+2)} \\ [\boldsymbol{W}_i]_{2\times(n_1+2)} \end{bmatrix} \tag{S53}$$

After recursive calculation up to the N-th layer, we obtain the total stiffness matrix $\boldsymbol{K}_{G_N}$. Using Eq. (37), this is converted into the global compliance matrix $\boldsymbol{S}_{G_N}$. We combine the boundary conditions for the incident wave amplitude vector $\boldsymbol{I}$ (assuming normalized amplitude of 1), reflected wave amplitude vector $\boldsymbol{R}$, and transmitted wave amplitude vector $\boldsymbol{T_r}$ to form the system:

$$\begin{cases} u_{z_0}(z_1) = B_0^{11} + RB_0^{12} \\ -P_{f_0}(z_1) = B_0^{21} + RB_0^{22} \end{cases} \quad \begin{cases} u_{z_{N+1}}(z_{N+1}) = T_r B_{N+1}^{11} \\ -P_{f_{N+1}}(z_{N+1}) = T_r B_{N+1}^{21} \end{cases} \tag{S54}$$

Based on the boundary conditions in Eq. (46) and the global compliance matrix, the transmission and reflection coefficients are solved as:

$$\begin{bmatrix} R \\ T_r \end{bmatrix} = \begin{bmatrix} B_0^{12} - S_{G_{N,2;2}} B_0^{22} & -S_{G_{N,2;4}} B_{N+1}^{21} \\ -S_{G_{N,4;2}} B_0^{22} & B_{N+1}^{11} - S_{G_{N,4;4}} B_{N+1}^{21} \end{bmatrix}^{-1} \begin{bmatrix} S_{G_{N,2;2}} B_0^{21} - B_0^{11} \\ S_{G_{N,4;2}} B_0^{21} \end{bmatrix} \tag{S55}$$

**Supplementary Table 1** Skull Parameters.

| Parameters | Value |
|---|---|
| Total Thickness | 3.87 mm |
| Outer Cortical Bone Thickness | 1.60 mm |
| Middle Diploe Bone Thickness | 0.68 mm |
| Inner Cortical Bone Thickness | 1.59 mm |
| Middle Diploe Bone Porosity | 78.1 % |
| Cortical Bone Longitudinal Velocity | 3500 m/s |
| Cortical Bone Transverse Velocity | 1460 m/s |
| Skull Density | 1800 kg/m$^3$ |
| Soft Tissue Velocity | 1500 m/s |
| Soft Tissue Density | 1000 kg/m$^3$ |

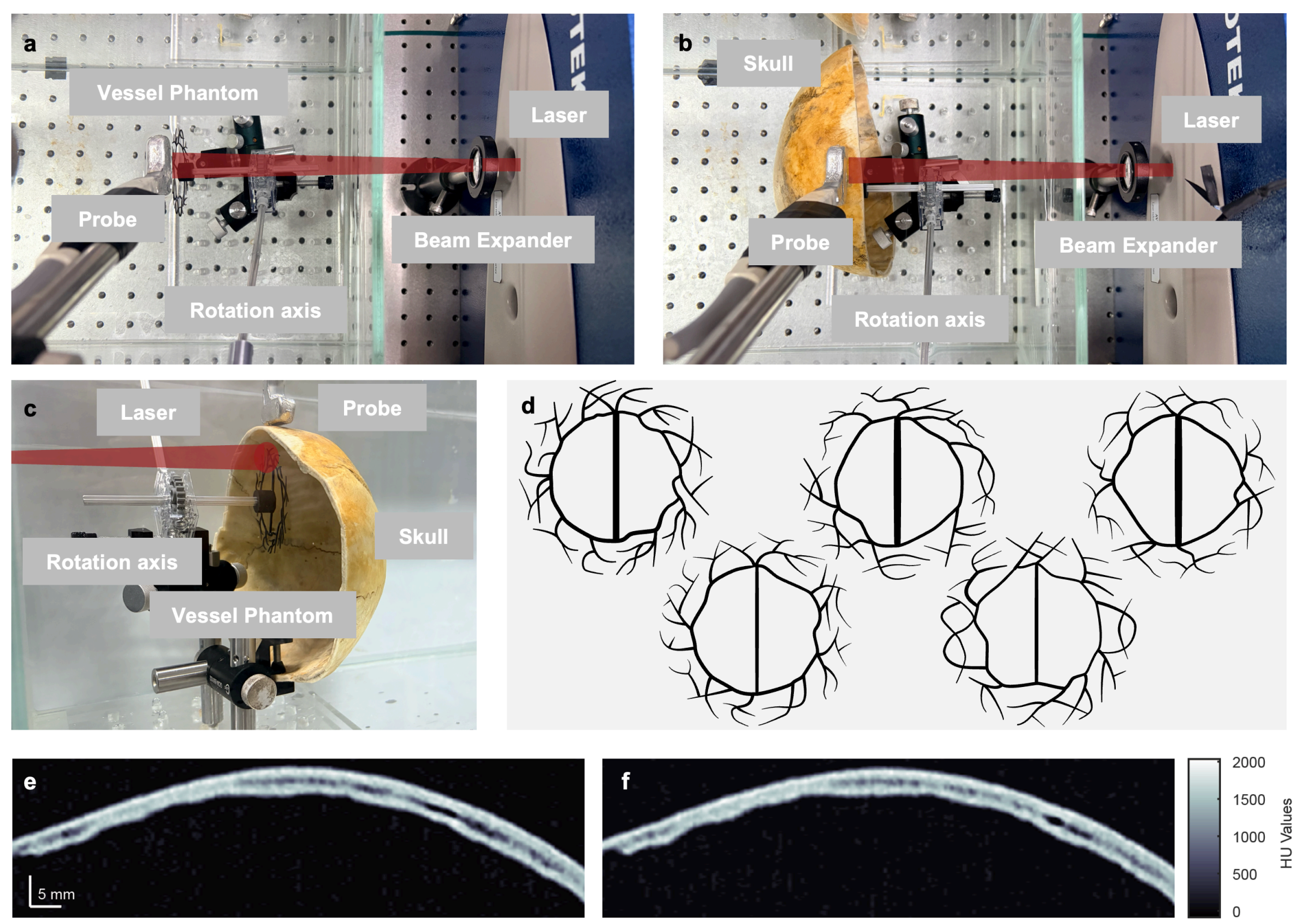

**Supplementary Fig. 2** Experimental setup for tPAI. a, Photograph of the setup without the skull, used to acquire GT images of the vessel phantom. b, c, Photographs of the complete experimental setup, including the ex vivo skull, from two different perspectives. d, The 3D-printed vessel phantom structures. e, CT image of the trained position. f, CT image of the cross position.

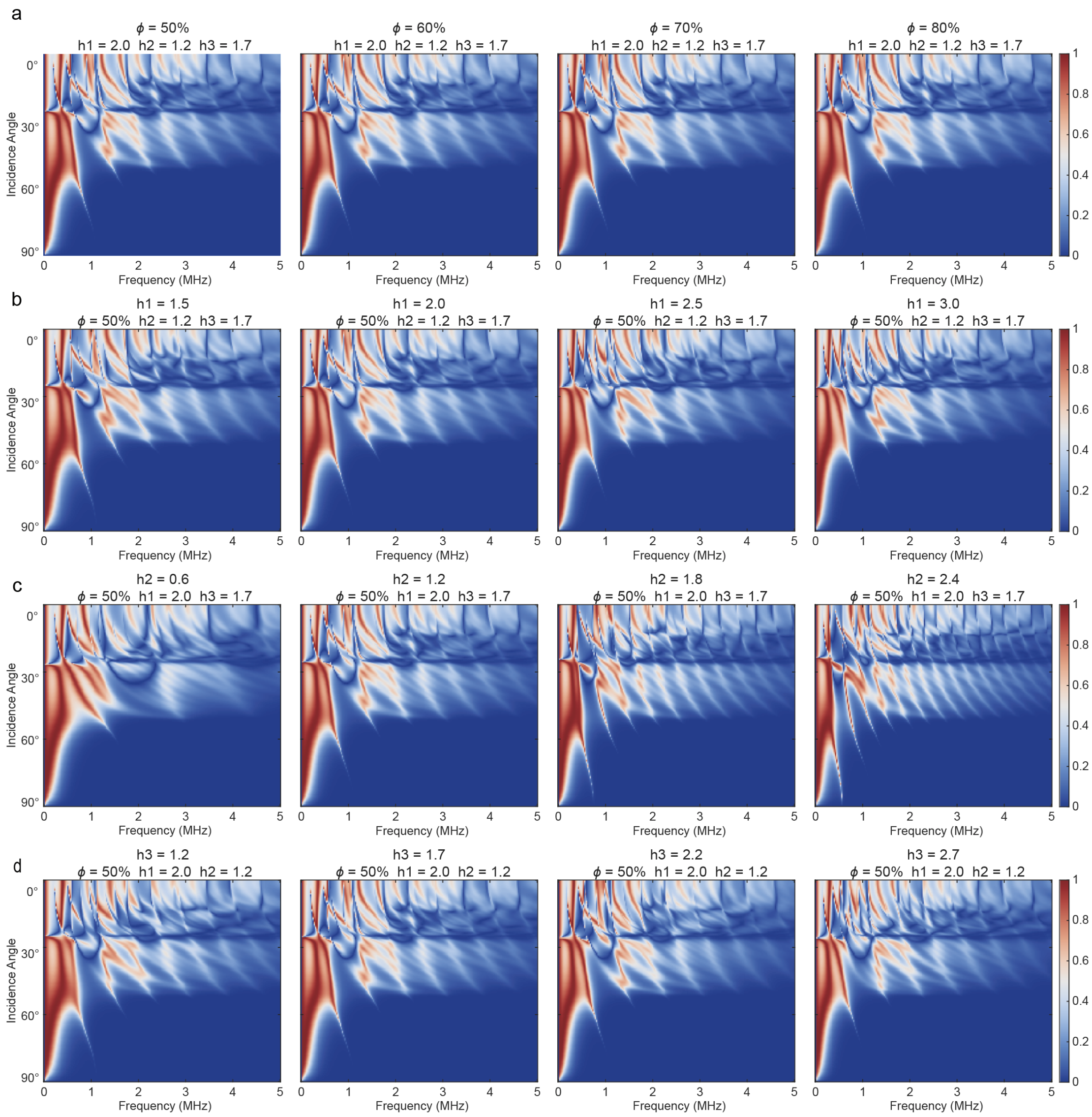

**Supplementary Fig. 3** Transmittance coefficient map as a function of frequency (0–5 MHz) and incidence angle (0°–90°) under variations of: (a) middle diploe layer porosity ($\phi = 50\% - 80\%$); (b) inner cortical layer thickness ($h_1 = 1.5 - 3.0\ mm$); (c) middle diploe layer thickness ($h_2 = 0.6 - 2.4\ mm$); and (d) outer cortical layer thickness ($h_3 = 0.6 - 2.4\ mm$).

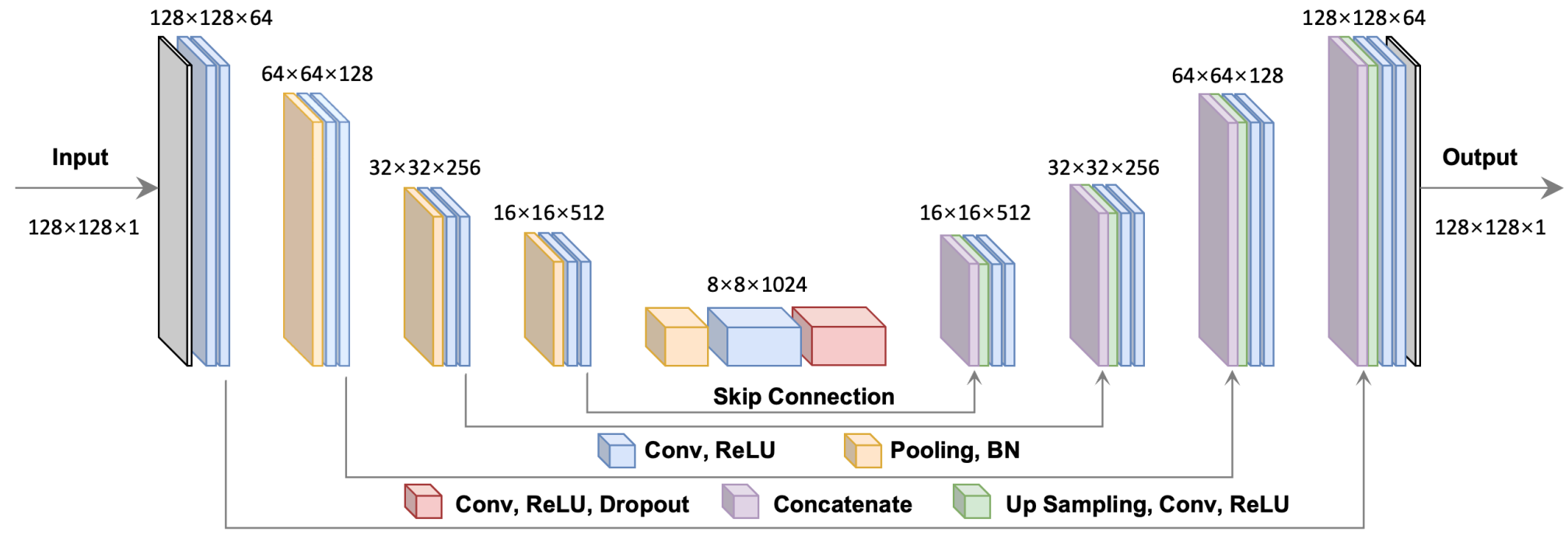

**Supplementary Fig. 4** Proposed PCL-CMM model based on the modified U-Net architecture. The numbers above each block indicate the dimensions of the feature maps in the format Height × Width × Channels.